\pdfoutput=1

\documentclass[a4paper,cits]{JINST}
\usepackage{comment}

\title{NEXT-DEMO: a prototype for the NEXT experiment}

\author{Paola Ferrario$^a$ on behalf of the NEXT Collaboration\\
\llap{$^a$} Instituto de F\'isica Corpuscular (Universidad de Valencia - CSIC),\\
  c/ Catedr\'atico Jos\'e Beltr\'an 2,  E-46980 Valencia,  Spain\\
  E-mail: \email{paola.ferrario@ific.uv.es}}
  
\abstract{NEXT is a high-pressure (10-15 bar) xenon Time Projection Chamber dedicated to neutrinoless double beta decay searches.
NEXT-DEMO is a large prototype of the NEXT experiment, constructed to demonstrate the feasibility of the NEXT detector concept, in particular the capability of the TPC to achieve an optimal energy resolution at the Xe $Q_{\beta\beta}$ energy (2.458 MeV). It has been operating at IFIC, in Valencia, for one year and a half and a number of results has been achieved. Recently, the reflector panels of the prototype have been coated with a wavelength shifter (TPB), in order to improve light collection. We have obtained an energy resolution of 2.3\% FWHM at 511 keV, using  Na-22 radioactive sources, corresponding to 1\% FWHM once extrapolated to $Q_{\beta\beta}$. 
}


\begin{document}

\section{Introduction}\label{sec:Introduction}

NEXT is a neutrino experiment which will search for neutrinoless double beta decay ($\beta\beta 0\nu$) in xenon enriched in isotope Xe-136 \cite{GomezCadenas:2012jv}. 

While the two-neutrino mode of the double beta decay has already been measured in a number of isotopes, the zero-neutrino mode remains unobserved and holds important implications regarding the nature of the neutrino. All the other fermions in the Standard Model have a Dirac-type mass, but the neutrino, being chargeless, can have a Majorana-type mass, and be its own antiparticle. Majorana neutrinos provide an attractive explanation for the smallness of neutrino masses, via the so-called seesaw mechanism. Besides, a Majorana nature of neutrinos would imply a violation of the lepton number. Lepton number violation is one of the basic ingredients, together with CP violation, to explain the asymmetry between matter and antimatter in our universe. If the neutrino is the same as its own antiparticle, the neutrinoless double beta nuclear decay is allowed, since the two emitted antineutrinos can annihilate and only two electrons are emitted. The reverse is equally true, namely that the existence of such a decay implies the Majorana nature of neutrinos. Thus, the observation of neutrinoless double beta decay would be a major discovery in particle physics \cite{Avignone:2007fu, Bilenky:2010zz}.

The signal of a neutrinoless double beta decay is a peak in the kinetic energy spectrum of the outcoming electrons ($Q_{\beta\beta}$). Thus, the most important characteristics of a $\beta\beta 0\nu$ experiment are optimal energy resolution, to constrain the region of interest as much as possible, and high background rejection, to get rid of the spurious events with almost the same energy as the $Q_{\beta\beta}$.

The race to the discovery of neutrinoless double beta decay is under way, with a number of experiments already running. They exploit different isotopes and experimental techniques each having advantages and disadvantages (for a review on $\beta\beta 0\nu$ experiments, see, for instance, ref. \cite{Elliott:2012sp}).

NEXT is a high-pressure,  xenon electroluminescent Time Projection Chamber (TPC) \cite{Lepeltier:2007, Nygren:2009zz}, with separated planes for energy measurement and for tracking \cite{Granena:2009it,:2012ha}. It will be placed in the Canfranc Underground Laboratory, in the Spanish Pyrenees. Its novel concept meets the essential requirements of a $\beta\beta 0\nu$ experiment mentioned above. On the side of the energy resolution, the Fano factor of gaseous xenon is much lower than the one of liquid xenon. Moreover, electroluminescence, used to obtain signal amplification, has very low fluctuations in gain.  On the other side, in xenon at 10 bar, it is possible to take advantage of the topological signature of the event: the two electrons of $\beta\beta 0\nu$ leave a track of about 15 cm, with almost constant energy deposition and two big ``blobs'' of energy at the ends, caused by the Bragg-peaks of the stopping electrons. This signature is an excellent handle for background rejection.

From 2009 to 2012 an intense R\&D program has been carried out by the Collaboration. The program has culminated in the construction, commissioning and operation of the NEXT-DEMO prototype, located at IFIC, in Valencia. In addition, a smaller prototype, operating at LBNL, has been used to perform detailed energy resolution studies.

\section{NEXT-DEMO}\label{sec:nextdemo}

\begin{figure}
  \begin{center}
\includegraphics[width=.65\textwidth]{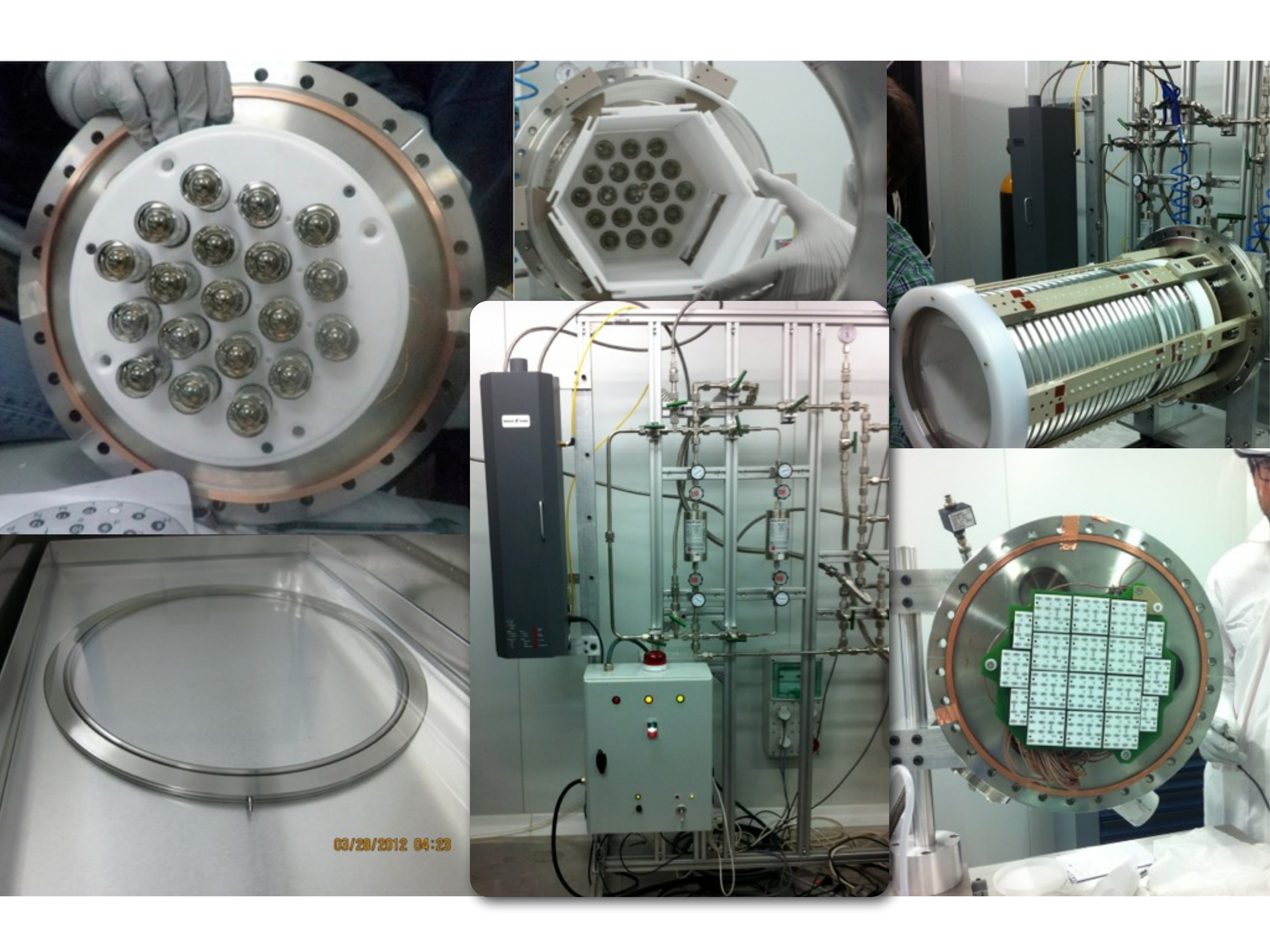}
\caption{Components of the NEXT-DEMO detector. From the top left, clockwise: the PMT plane, the light tube with reflector panels, the field cage, the tracking plane with MPPCs, the gas system with the hot getter and the EL meshes.}
\label{fig:collage}
\end{center}
\end{figure}

NEXT-DEMO is a large scale (1:10) prototype with the aim of testing the instrumental concept of NEXT: it is an asymmetric TPC of 60~cm of length and 30 cm of diameter, which can hold up to 15 bar of xenon gas. The maximum capacity of the detector is 10 kg but in its current configuration (the fiducial volume is a hexagon of 16 cm diameter and 30 cm length) it holds 4 kg of natural (non-enriched) xenon at 10 bar. On the cathode side a matrix of 19 Hamamatsu R7378A photomultipliers (PMT) measures the energy of each event, while, on the anode, $\sim$ 300 Hamamatsu silicon photomultipliers (MPPCs) with $\sim$ 1 cm of pixelization provide its position in the perpendicular plane. Combined with the time measurement of the event, a full three-dimensional reconstruction is provided.  The main components of the detector are shown in figure \ref{fig:collage}.

\begin{figure}
\includegraphics[width=1\textwidth]{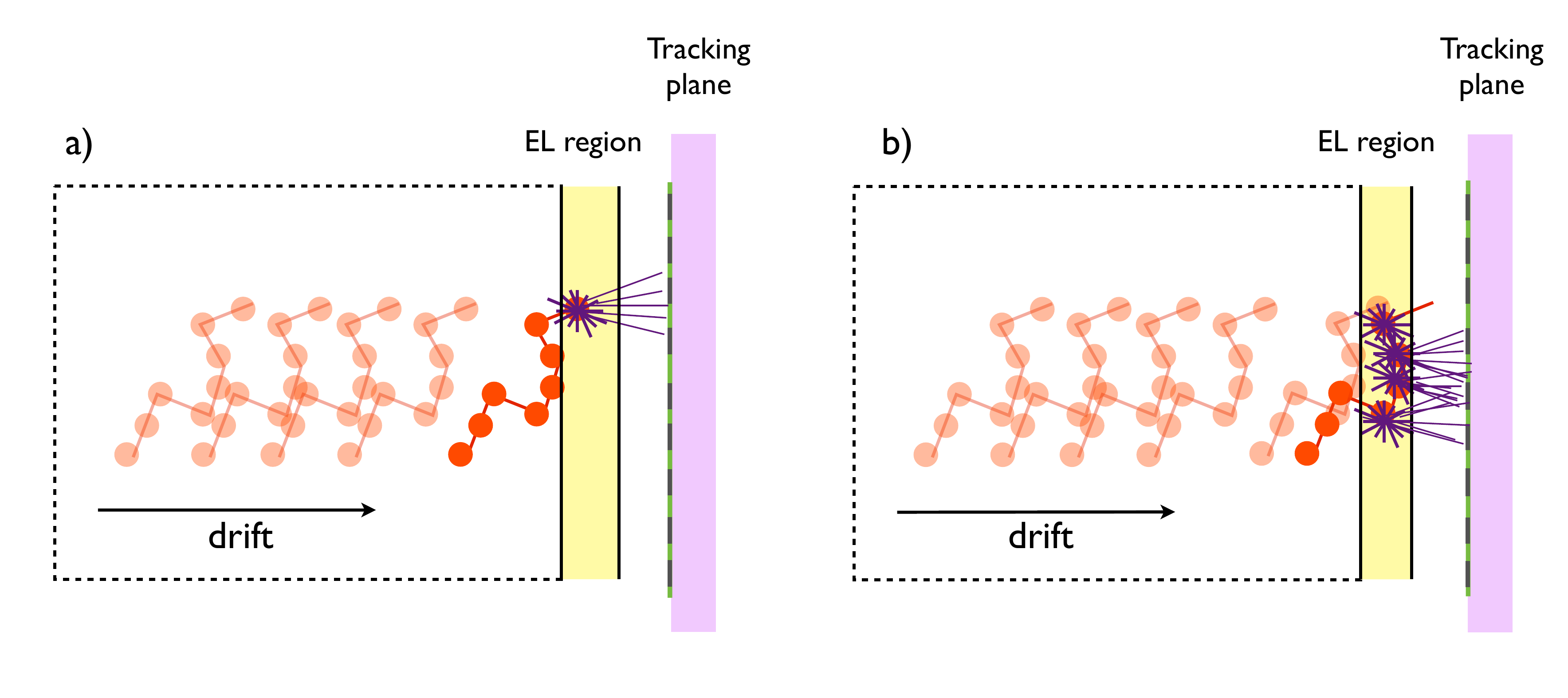}
\caption{Formation of the bidimensional image on the tracking plane: a) the first arriving charge at the EL region, b) the arriving charge at a later stage. }
\label{fig:elscheme}
\end{figure}

When a charged particle enters the gas, it deposits its energy through both scintillation and ionization of the gas molecules. The scintillation light (UV photons at $\sim$ 172 nm) is registered by the PMTs on the cathode side and gives the starting time of the event. The ionization electrons are drifted by an electric field of around 500 V/cm all the way through the drift region until they enter a region of moderately higher field where they are accelerated and secondary scintillation (but not ionization) occurs. This process, called electroluminescence (EL), results in an amplification of the signal, which grows linearly with the electric field as long as its magnitude remains below the ionization threshold. The PMTs in the cathode detect the EL light, giving a precise measurement of the energy of the event. In the anode, the distribution of the EL light on the silicon photomultipliers is, in every moment, a 2D picture of the track at a given position along the axis, as shown in figure \ref{fig:elscheme}. Knowing the starting time of the event, the absolute position along the TPC axis can also be reconstructed.

NEXT-DEMO has been operating successfully for more than one year and has demonstrated: a) operation stability, with no leaks and very few sparks, even at high voltages; (b) energy resolution; (c) a good electron drift lifetime, of the order of several milliseconds.  In the following section the main results will be presented.


\section{Results and analysis}\label{sec:results}

NEXT-DEMO started running in November 2010, proving an excellent operation through the collection of millions of events. Radioactive sources such as Cs-137 and Na-22 have been extensively used for energy calibration. 
During the Na-22 campaign, the source was positioned in a port on one side of the chamber, quite close to the cathode. A NaI scintillator was placed behind the source in order to tag one of the two back-to-back  511 keV gammas coming from the annihilation of the beta positron. A digital trigger was used, providing a variety of on-line triggering methods (coincidence with the NaI signal, primary scintillation, EL signal or a combination of both). 

There are two important effects that spoil the energy resolution and that have been corrected during the analysis of data. One is the electron trapping along the drift, which produces a loss of ionization electrons, thus a loss of electroluminescent light. The other one is the radial dependence of the collection of the EL light by the PMTs in the cathode: the farther from the axis the event takes place, the fewer the number of photons detected in the PMTs.

The first part of data collection has been carried through with the original reflector panels, while, in a second stage, those panels were coated with a wavelength shifter, in order to improve the reflection of light and its collection in the cathode. A thin film of the organic chemical compound Tetraphenyl butadiene (TPB) has been applied on the sides of the panels that face the drift region. TPB absorbs the ultraviolet light of xenon scintillation and has an emission peak in the blue region, at around 430 nm \cite{Lally:1995bb,Gehman:2011xm,Boccone:2009kk}. Reflector panels are more efficient at those wavelengths and the quantum efficiency of the PMTs is approximately a factor of two higher. This improvement results in an increase in light collection by a factor of 3--5. 

\begin{figure}
  \begin{center}
\includegraphics[width=.75\textwidth]{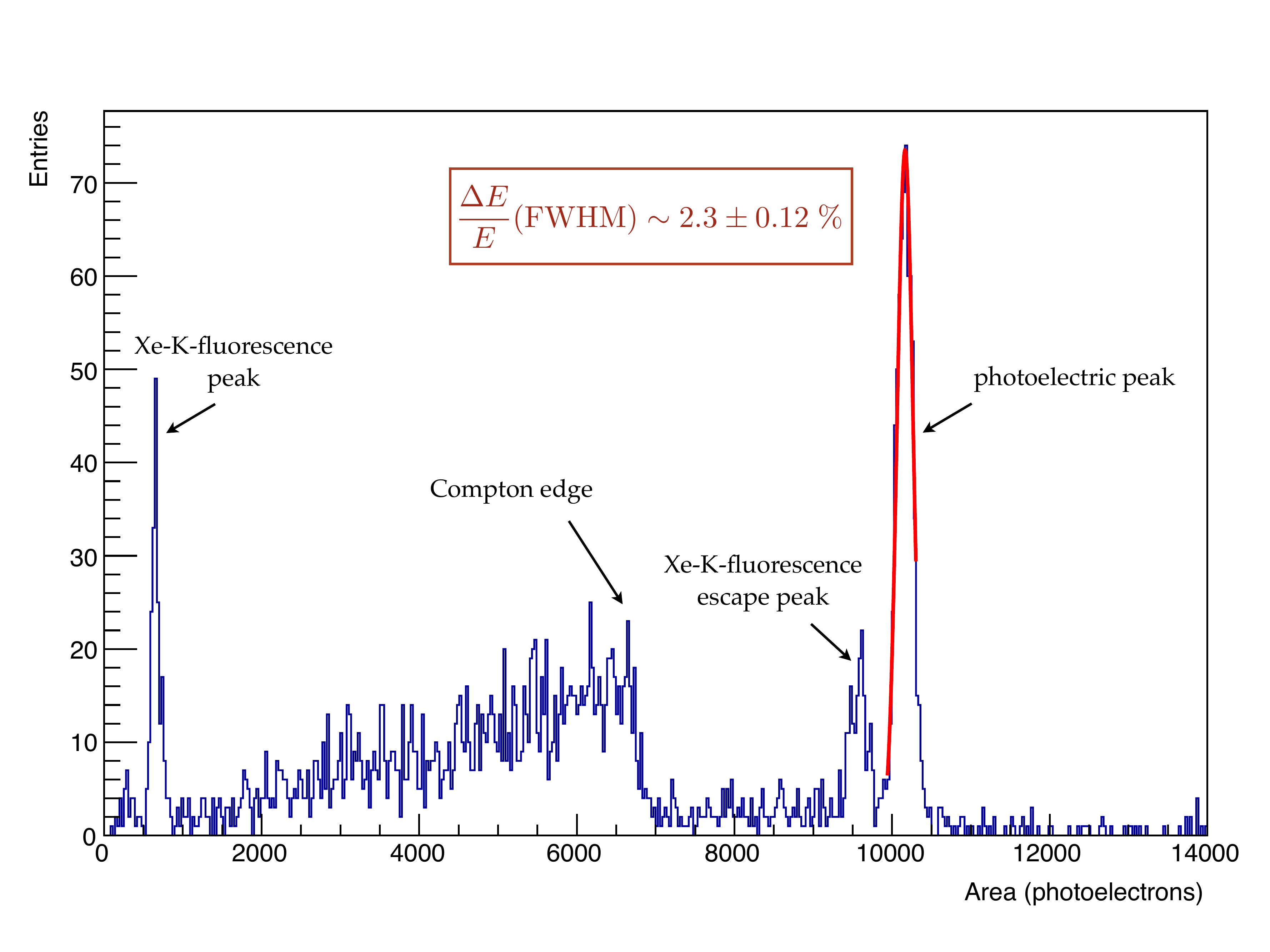}
    \end{center}
\caption{Typical pulse height distribution for 511 keV gamma interactions. An energy resolution of 2.3 \% FWHM is obtained at 511 keV, after attachment and radial corrections. It extrapolates to 1\% FWHM at $Q_{\beta\beta}$. The energy axis is displayed in number of photoelectrons.}
\label{fig:resolution}
\end{figure}

In figure \ref{fig:resolution} we present a typical pulse height distribution obtained for the 511 keV interactions  in the detector, after implementing the attachment and radial corrections. We find an energy resolution of 2.2\% FWHM for the photoelectric peak, for both cases with and without TPB coating. This value extrapolates to 1\% FWHM at the $Q_{\beta\beta}$ (2.458 MeV). The lack of further improvement with the better collection of light is probably due to intrinsic limitations of the chamber and calibration effects. It is the first time that such a resolution is reached in a high-pressure xenon TPC.  This result proves the better feature of gaseous compared to liquid xenon (see, for instance, the $\sim 4\%$ FWHM at the Xe $Q_{\beta\beta}$ of the EXO-200 experiment \cite{Auger:2012ar}).



In the last phase of the Na-22 campaign the PMT plane for the tracking has been replaced by an array of MPPCs \cite{Renker:2009zz, McConkey:2011zz,Lightfoot:2008ig,:2012za}. Since MPPCs are not sensitive to the UV light, they also need to be coated with TPB. This produces an overall improvement of light collection, since TPB coating emits the shifted light isotropically. The data collection and analysis of this third phase is still ongoing, but the first results of EL light detection by means of MPPCs have already been recorded.

\section{Conclusions and outlook}\label{sec:conclusions}
A large, 15 bar Xe TPC based on primary and secondary scintillation readout was built and has been in operation for more than one year. It has allowed us to develop the needed know-how to build the final detector for NEXT and has shown the validity of all the proposed solutions. The experimental results demonstrate that such a detector can achieve an unprecedented energy resolution of 2.3\% FWHM at 511 keV, which corresponds to 1\% FWHM at 2.458 MeV. Therefore, the NEXT concept presents a clear advantage for the detection of neutrinoless double beta decay, when compared to the different detector technologies available or in operation in similar experiments. The next step in the NEXT-DEMO operation is the study of the tracking capability of MPPCs: a new tracking plane has been installed and data taking and analysis are ongoing. Meanwhile, the construction of the NEXT final detector has started in Canfranc: great times are approaching...
\acknowledgments

The NEXT Collaboration acknowledges support from the following agencies and institutions: the Spanish Ministerio de Econom\'ia y Competitividad under grants CONSOLIDER-Ingenio 2010 CSD2008-0037 (CUP), FPA2009-13697-C04-04 and RYC-2008-03169;  the European Commission under the European Research Council Starting Grant ERC-2009-StG-240054 (T-REX) of the IDEAS program of the 7th EU Framework Program; and the Director, Office of Science, Office of Basic Energy Sciences, of the US Department of Energy under contract no. DE-AC02-05CH11231.  We also acknowledge support from the Portuguese FCT and FEDER through program COMPETE, projects PTDC/FIS/103860/2008 and PTDC/FIS/112272/2009. J. Renner (LBNL) acknowledges the support of a US DOE NNSA Stewardship Science Graduate Fellowship under contract no. DE-FC52-08NA28752.

\bibliographystyle{JHEP}
\bibliography{biblio}
\end{document}